\documentclass[10pt,aps,pra,reprint,showpacs,superscriptaddress]{revtex4-1}
\usepackage[english]{babel}
\usepackage[T1]{fontenc}
\usepackage{natbib}
\usepackage{soul}
\usepackage{amssymb}
\usepackage{bm}
\usepackage{amsmath}
\usepackage[mathcal]{eucal}
\usepackage{color}
\usepackage{graphicx}
\usepackage{hyperref}
\usepackage{mathtools}
\usepackage{url}
\usepackage{tikz}
\usetikzlibrary{shapes}
\usetikzlibrary{decorations}
\usetikzlibrary{arrows}
\usetikzlibrary{matrix}

\newcommand {\nc} {\newcommand}
\nc {\beq} {\begin{eqnarray}}
\nc {\eeqn} [1] {\label{#1} \end{eqnarray}}
\nc {\eoln} [1] {\label{#1} \\}
\nc {\eol} {\nonumber \\}
\nc {\rref} [1] {(\ref{#1})}
\nc {\Eq} [1] {Eq.~(\ref{#1})}
\nc {\Ref} [1] {Ref.~\cite{#1}}
\nc {\la} {\mbox{$\langle$}}
\nc {\ra} {\mbox{$\rangle$}}
\nc {\dem} {\mbox{$\frac{1}{2}$}}
\nc {\cP} {\mathcal{P}}
\nc {\cN} {\mathcal{N}}
\nc {\ve} [1] {\mbox{\boldmath $#1$}}
\nc {\arrow} [2] {\mbox{$\mathop{\rightarrow}\limits_{#1 \rightarrow #2}$}}
\nc {\red}[1] {\textcolor{red}{#1}}
\nc {\mc}[3] {\multicolumn{#1}{#2}{#3}}
\nc {\dd}{\, \mathrm{d}}
\nc {\bs}[1]{\boldsymbol{#1}}
\nc {\ket}[1]{\vert #1 \rangle}
\nc {\bra}[1]{\langle #1 \vert}
\nc {\abs}[1]{\vert #1 \vert}
\nc {\avg}[1]{\langle #1 \rangle}
\nc {\braket}[2]{\langle #1 \vphantom{#2} \vert #2 \vphantom{#1} \rangle}
\nc {\abss}[1]{\left| #1 \right|}

\begin{document}

\title{Multiconfiguration calculations of electronic isotope shift factors in~Zn~\textsc{i}}
\author{Livio Filippin}
\email[]{Livio.Filippin@ulb.ac.be}
\affiliation{Chimie Quantique et Photophysique, Universit\'{e} libre de Bruxelles, B-1050 Brussels, Belgium}
\author{Jacek Biero{\'n}}
\email[]{jacek.bieron@uj.edu.pl}
\affiliation{Instytut Fizyki imienia~Mariana~Smoluchowskiego, Uniwersytet Jagiello{\'n}ski, PL-30-348 Krak{\'o}w, Poland}
\author{Gediminas Gaigalas}
\email[]{Gediminas.Gaigalas@tfai.vu.lt}
\affiliation{Institute of Theoretical Physics and Astronomy, Vilnius University, LT-10222 Vilnius, Lithuania}
\author{Michel Godefroid}
\email[]{mrgodef@ulb.ac.be}
\affiliation{Chimie Quantique et Photophysique, Universit\'{e} libre de Bruxelles, B-1050 Brussels, Belgium}
\author{Per J\"{o}nsson}
\email[]{per.jonsson@mah.se}
\affiliation{Group for Materials Science and Applied Mathematics, Malm\"{o} University, S-20506 Malm\"{o}, Sweden}

\date{\today}

\begin{abstract}
The present work reports results from systematic multiconfiguration Dirac-Hartree-Fock calculations of electronic isotope shift factors for a set of transitions between low-lying states in neutral zinc. These electronic quantities together with observed isotope shifts between different pairs of isotopes provide the changes in mean-square charge radii of the atomic nuclei. Within this computational approach, different models for electron correlation are explored in a systematic way to determine a reliable computational strategy and to estimate theoretical error bars of the isotope shift factors.
\end{abstract}


\maketitle


\section{Introduction}
\label{sec:intro}

When the effects of the finite mass and the extended spatial charge distribution of the nucleus are taken into account in a Hamiltonian describing an atomic system, the electronic energy levels undergo a small, isotope-dependent shift~\cite{NGG13}. The isotope shift (IS) of spectral lines, which consists of the mass shift (MS) and the field shift (FS), plays a key role in extracting the changes in mean-square charge radii of the atomic nuclei~\cite{Ki84,CCF12,NLG15}. For a given atomic transition $k$ with frequency~$\nu_k$, it is assumed that the electronic response of the atom to variations of the nuclear mass and charge distribution can be described by only two factors: the mass-shift factor $\Delta K_{k,\text{MS}}$ and the field-shift factor $F_k$. The observed IS $\delta \nu_k^{A,A'}$ between any pair of isotopes with mass numbers $A$ and $A'$ is related to the difference in nuclear masses and in mean-square charge radii, $\delta \langle r^2 \rangle^{A,A'}$~\cite{NGG13,Ki84}.

This work focuses on two transitions between low-lying levels of neutral zinc (Zn~\textsc{i}), the lightest element of group 12 (IIB), that have been under investigation in laser spectroscopy experiments along the Zn isotopic chain. Campbell \textit{et al.}~\cite{CBG97} measured the isotope shifts between stable isotopes ($^{64,66-68,70}$Zn) for the $4s^{2}~^{1}S_{0} \rightarrow 4s4p~^{3}P^{o}_{1}$ (307.6~nm) transition using a crossed atomic-laser beam experiment. Specific mass shifts (SMSs) were extracted, and a large value has been assigned to the ground state, emphasizing the substantial $3d$ core-valence polarization. Recently, Yang \textit{et al.}~\cite{YWX16} investigated the $4s4p~^{3}P^{o}_{2} \rightarrow 4s5s~^{3}S_{1}$ (481.2~nm) transition in a bunched-beam collinear laser spectroscopy experiment to determine nuclear properties of the $^{79}$Zn isotope. The isomer shift between the nuclear ground state and the long-lived $1/2^+$ isomeric state was measured, and the change of the mean-square charge radii of $^{79,79m}$Zn has been extracted via the MS and FS electronic factors. The latter were obtained from a King-plot process~\cite{Ki84} using the root-mean-square charge radii of isotopes from Refs.~\cite{CBG97,FH04}.

There are many theoretical studies of properties such as oscillator strengths, lifetimes, polarizabilities and hyperfine structure constants in Zn~\textsc{i} and Zn-like ions~\cite{FH78,FH79,BG80,Hi89,BF92,CCH94,FH95,GM03,MH05,GM06,JAS06,LHZ06,FZ07,BJS08,ALC08,CC10,CC10b,SS10,LGZ11,CC14}. By contrast, to the best of our knowledge, no recent paper reporting on theoretical IS electronic factors in Zn~\textsc{i} has been published since the pioneer works led by Bauche and Crubellier~\cite{BC70,BC74} reporting only on SMS factors, and by Blundell \textit{et al.}~\cite{BBP85,BBP87} only on FS factors. Hence, we reinvestigate the two above-cited transitions in Zn~\textsc{i} by performing \textit{ab initio} calculations of IS electronic factors using the multiconfiguration Dirac-Hartree-Fock (MCDHF) method implemented in the \textsc{ris\small{3}}/\textsc{grasp\small{2}k} program package~\cite{NGG13,JGB13}. Using the MCDHF method, the computational scheme is based on the estimation of the expectation values of the one- and two-body recoil Hamiltonian for a given isotope, including relativistic corrections derived by Shabaev~\cite{Sh85,Sh88}, combined with the calculation of the total electron densities at the origin.

This approach has recently been performed on neutral copper (Cu~\textsc{i})~\cite{BCF16,CG16} to determine a set of $\delta \langle r^2 \rangle^{65,A'}$ values from the corresponding observed IS. Later on, it has been applied to neutral magnesium (Mg~\textsc{i})~\cite{FGE16} and neutral aluminium (Al~\textsc{i})~\cite{FBE16}, where IS factors have been computed for transitions between low-lying states. In the present work, different electron correlation models are applied to Zn~\textsc{i} to estimate theoretical error bars of the IS factors.

In Sec.~\ref{sec:method}, the principles of the MCDHF method are summarized. In Sec.~\ref{sec:isotope_shifts}, the relativistic expressions of the MS and FS factors are recalled. Section~\ref{sec:AS} presents the active space expansion strategies adopted for the electron correlation models. In Sec.~\ref{sec:results}, numerical results of the MS and FS factors are reported for each of the two studied transitions in Zn~\textsc{i}. Section~\ref{sec:conc} reports conclusions.


\section{Numerical method}
\label{sec:method}

The MCDHF method~\cite{Gr07}, as implemented in the \textsc{grasp{\small 2}k} program package~\cite{JGB13,JHF07}, is the fully relativistic counterpart of the non-relativistic multiconfiguration Hartree-Fock (MCHF) method~\cite{FTG07,FGB16}. The MCDHF method is employed to obtain wave functions that are referred to as atomic state functions (ASF), i.e., approximate eigenfunctions of the Dirac-Coulomb Hamiltonian given by
\beq
\mathcal{H}_{\text{DC}} = \sum_{i=1}^N [c \, \bs{\alpha}_i \cdot \bs{p}_i + (\beta_i - 1)c^2 + V_{\text{nuc}}(r_i)] + \sum_{i<j}^N \frac{1}{r_{ij}}, \eol
\eeqn{eq_DC_Hamiltonian}
where $V_{\text{nuc}}(r_i)$ is the nuclear potential corresponding to an extended nuclear charge distribution function, $c$ is the speed of light and $\bs{\alpha}$ and $\beta$ are the $(4 \times 4)$ Dirac matrices. An ASF, $\Psi(\gamma\,\Pi JM_J)$, is given as an expansion over $N_{\text{CSFs}}$ $jj$-coupled configuration state functions (CSFs), $\Phi(\gamma_{\nu}\Pi JM_J)$, with the same parity $\Pi$, total angular momentum $J$ and its projection on the $z$-axis, $M_J$:
\beq
\vert \Psi(\gamma\,\Pi JM_J) \rangle = \sum_{\nu=1}^{N_{\text{CSFs}}} c_{\nu} \, \vert \Phi(\gamma_{\nu}\,\Pi JM_J) \rangle.
\eeqn{eq_ASF}

In the MCDHF method, the one-electron radial functions used to construct the CSFs and the expansion coefficients $c_{\nu}$ are determined variationally so as to leave the energy functional
\beq
E = \sum_{\mu,\nu}^{N_{\text{CSFs}}} c_{\mu} c_{\nu} \langle \Phi(\gamma_{\mu}\,\Pi JM_J) \vert \mathcal{H}_{\text{DC}} \vert \Phi(\gamma_{\nu}\,\Pi JM_J) \rangle
\eeqn{eq_energy_functional}
and additional terms for preserving the orthonormality of the radial orbitals stationary with respect to their variations. The resulting coupled radial equations are solved iteratively in the self-consistent field (SCF) procedure. Once radial functions have been determined, a configuration-interaction (CI) diagonalization of Hamiltonian~\rref{eq_DC_Hamiltonian} is performed over the set of configuration states, providing the expansion coefficients for building the potentials for the next iteration. The SCF and CI coupled processes are repeated until convergence of the total wave function~\rref{eq_ASF} and energy~\rref{eq_energy_functional} is reached.


\section{Isotope shift theory}
\label{sec:isotope_shifts}

The finite mass of the nucleus gives rise to a recoil effect that shifts the level energies slightly, called the mass shift (MS). Due to the variation of the IS between the upper and lower levels, the transition IS arises as a difference between the IS for the two levels. Furthermore, the transition frequency MS between two isotopes, $A$ and $A'$, with nuclear masses $M$ and $M'$, is written as the sum of normal mass shift (NMS) and specific mass shift (SMS),
\beq
\delta \nu_{k,\text{MS}}^{A,A'} \equiv \nu_{k,\text{MS}}^{A} - \nu_{k,\text{MS}}^{A'} = \delta \nu_{k,\text{NMS}}^{A,A'} + \delta \nu_{k,\text{SMS}}^{A,A'},
\eeqn{eq_delta_nu_MS}
and can be expressed in terms of a single parameter
\beq
\delta \nu_{k,\text{MS}}^{A,A'} = \left( \frac{1}{M} - \frac{1}{M'} \right) \frac{\Delta K_{k,\text{MS}}}{h} = \left( \frac{1}{M} - \frac{1}{M'} \right) \Delta \tilde{K}_{k,\text{MS}}. \eol
\eeqn{eq_delta_nu_MS_2}
Here, the mass shift factor $\Delta K_{k,\text{MS}}=(K_{\text{MS}}^u-K_{\text{MS}}^l)$ is the difference of the $K_{\text{MS}}=K_{\text{NMS}}+K_{\text{SMS}}$ factors of the upper ($u$) and lower ($l$) levels involved in the transition $k$. For the $\Delta \tilde{K}$ factors, the unit (GHz~u) is often used in the literature. As far as conversion factors are concerned, we use $\Delta K_{k,\text{MS}}\,[m_eE_{\text{h}}]=3609.4824\,\Delta \tilde{K}_{k,\text{MS}}\,[\text{GHz~u}]$.

Neglecting terms of higher order than $\delta \langle r^2 \rangle$ in the Seltzer moment (or nuclear factor)~\cite{Se69}
\beq
\lambda^{A,A'} = \delta \langle r^2 \rangle^{A,A'} + b_1 \delta \langle r^4 \rangle^{A,A'} + b_2 \delta \langle r^6 \rangle^{A,A'} + \cdots, \eol
\eeqn{eq_Seltzer_moment}
the line frequency shift in the transition $k$ arising from the difference in nuclear charge distributions between two isotopes, $A$ and $A'$, can be written as~\cite{FBH95,TFR85,BBP87}
\beq
\delta \nu_{k,\text{FS}}^{A,A'} \equiv \nu_{k,\text{FS}}^{A} - \nu_{k,\text{FS}}^{A'} = F_k \, \delta \langle r^2 \rangle^{A,A'}.
\eeqn{eq_delta_nu_FS}
In the expression above $\delta \langle r^2 \rangle^{A,A'} \equiv \langle r^2 \rangle^{A}-\langle r^2 \rangle^{A'}$ and $F_{k}$ is the electronic factor. Although not used in the current work, it should be mentioned that there are computationally tractable methods to include higher order Seltzer moments in the expression for the transition frequency shift~\cite{EJG16,PCE16}.

The total transition frequency shift is obtained by merely adding the MS, \rref{eq_delta_nu_MS}, and FS, \rref{eq_delta_nu_FS}, contributions:
\beq
\delta \nu_k^{A,A'} & = & \delta \nu_{k,\text{NMS}}^{A,A'} + \delta \nu_{k,\text{SMS}}^{A,A'} + \, \delta \nu_{k,\text{FS}}^{A,A'} \eol
& = & \left( \frac{1}{M} - \frac{1}{M'} \right) \Delta \tilde{K}_{k,\text{MS}} + F_k \, \delta \langle r^2 \rangle^{A,A'}.
\eeqn{eq_IS}

In this approximation, it is sufficient to describe the total frequency shift between the two isotopes $A$ and $A'$ with only the two electronic parameters given by the mass shift factor $\Delta \tilde{K}_{k,\text{MS}}$ and the field shift factor $F_k$. Furthermore, they relate line frequency shifts to nuclear properties given by the change in mass and mean-square charge radius. Both factors can be calculated from atomic theory, which is the subject of this work.

The main ideas of the method that is applied to compute these quantities are outlined here. More details can be found in the works by Shabaev \cite{Sh85,Sh88} and Palmer \cite{Pa88}, who pioneered the theory of the relativistic mass shift used in the present work. Gaidamauskas \textit{et al.}~\cite{GNR11} derived the tensorial form of the relativistic recoil operator implemented in \textsc{ris{\small 3}}~\cite{NGG13} and its extension~\cite{EJG16}.

The nuclear recoil corrections within the $(\alpha Z)^4m_e^2/M$ approximation~\cite{Sh85,Sh88} are obtained by evaluating the expectation values of the one- and two-body recoil Hamiltonian for a given isotope,
\beq
\mathcal{H}_{\text{MS}} = \frac{1}{2M} \sum_{i,j}^{N} \left( \bs{p}_i \cdot \bs{p}_j - \frac{\alpha Z}{r_i} \left( \bs{\alpha}_i + \frac{(\bs{\alpha}_i \cdot \bs{r}_i) \bs{r}_i}{r_i^2} \right) \cdot \bs{p}_j \right). \eol
\eeqn{eq_H_MS}
Separating the one-body $(i=j)$ and two-body $(i\neq j)$ terms that, respectively, constitute the NMS and SMS contributions, the Hamiltonian \rref{eq_H_MS} can be written
\beq
\mathcal{H}_{\text{MS}}=\mathcal{H}_{\text{NMS}}+\mathcal{H}_{\text{SMS}}.
\eeqn{eq_H_NMS_SMS} 

The NMS and SMS mass-independent $K$ factors are defined by the following expressions:
\beq
K_{\text{NMS}} \equiv M \langle \Psi \vert \mathcal{H}_{\text{NMS}} \vert \Psi \rangle,
\eeqn{eq_K_NMS}
\beq
K_{\text{SMS}} \equiv M \langle \Psi \vert \mathcal{H}_{\text{SMS}} \vert \Psi \rangle.
\eeqn{eq_K_SMS}

Within this approach, the electronic factor $F_{k}$ for the transition $k$ is estimated by
\beq
F_k = \frac{Z}{3\hbar} \left( \frac{e^2}{4\pi \epsilon_0} \right) \Delta \vert \Psi(0) \vert_k^2,
\eeqn{eq_F_k}
which is proportional to the change of the total electron probability density at the origin between levels $l$ and $u$,
\beq
\Delta \vert \Psi(0) \vert_k^2 = \Delta \rho_k^e(\bs{0}) = \rho_u^e(\bs{0}) - \rho_l^e(\bs{0}).
\eeqn{eq_Delta_Psi_0}

As potential $V_{\text{nuc}}(r_i)$ of \Eq{eq_DC_Hamiltonian} is isotope-dependent, the radial functions vary from one isotope to another, which defines isotopic relaxation. However, the latter is very small and hence neglected along the isotopic chain. Thus, the wave function $\Psi$ is optimized for a specific isotope within this approach.


\section{Active space expansion}
\label{sec:AS}

To effectively capture electron correlation, CSFs of a particular symmetry $J$ and parity $\Pi$ are generated through substitutions within an active space (AS) of orbitals, consisting of orbitals occupied in the reference configurations and correlation orbitals. From hardware and software limitations, it is impossible to use complete AS wave functions that would include all CSFs with appropriate $J$ and $\Pi$ for a given orbital AS. Hence the CSF expansions have to be constrained ensuring that major correlation substitutions are accounted for~\cite{FGB16}.

Single (S), double (D) (and triple (T), see Sec.~\ref{subsec:SrDT-SS}) substitutions are performed on either a single-reference (SR) set or a multireference (MR) set, the latter containing the CSFs that have large expansion coefficients and account for the major correlation effects. These substitutions take into account valence-valence (VV) and core-valence (CV) correlations. While the VV correlation model only allows SD substitutions from valence orbitals, the VV+CV correlation model considers restricted substitutions from core and valence orbitals. The restriction is applied to double (and triple) substitutions, denoted as SrD(T), in such a way that only one electron is substituted from the core shells, the other one (or two) has (have) to be substituted from the valence shells.

Zn~\textsc{i} has two valence electrons ($n=4$) outside an [Ar]$3d^{10}$ core. The MR sets (see Sec.~\ref{subsec:SrD-MR}) are obtained by performing SrDT substitutions from the $3d$ and the occupied valence orbitals to the $n=4$ valence orbitals + $5s/\{5s,5p\}/\{5s,6s\}$, depending on the targeted state $4s^{2}~^{1}S_{0}/4s4p~^{3}P^{o}_{1,2}/4s5s~^{3}S_{1}$ (maximum of one hole in the $3d$ orbital). An SCF procedure is then applied to the resulting CSFs, providing the orbital set and the expansion coefficients. Due to limited computer resources, such an MR set would be too large for subsequent calculations. Hence, only the CSFs whose expansion coefficients are, in absolute value, larger than a given MR cutoff are kept, i.e., $\vert c_\nu \vert >\varepsilon_{\text{MR}}$. The $\varepsilon_{\text{MR}}$ values and the resulting MR sets are listed in \tablename{~\ref{table_MR_composition}} for both transitions.

The $1s$ orbital is kept closed in all calculations, i.e., no substitutions from this orbital are allowed. Tests show that opening the $1s$ orbital does not affect the MS and FS factors within the accuracy attainable in the present calculations. Only orbitals occupied in the single configuration DHF approximation are treated as spectroscopic, i.e., are required to have a node structure similar to the corresponding hydrogenic orbitals~\cite{FGB16}. The occupied reference orbitals are frozen in all subsequent calculations. A \textit{layer} is defined as a subset of virtual orbitals with different angular symmetries, optimized simultaneously in one step, and frozen in all subsequent ones~\cite{FGB16}. One layer of $\{s, p, d, f, g\}$ symmetries and four of $\{s, p, d, f, g, h\}$ are successively generated. At each generation step, only the orbitals of the last layer are variational in the SCF procedure, all previously generated layers being kept frozen.

The effect of adding the Breit interaction to the Dirac-Coulomb Hamiltonian, \rref{eq_DC_Hamiltonian}, is found to be much smaller than the uncertainty in the transition IS factors with respect to the correlation model. This interaction has therefore been neglected in the procedure.

Within the three following correlation models, \textit{separate} orbital basis sets are optimized for the lower state and the upper state of each studied transition. For each state, the optimization procedures are summarised as follows:

\vspace{-0.33cm}

\subsection{SrD-SR model}
\label{subsec:SrD-SR}

(1) Perform a calculation using an SR set consisting of CSF(s) with the form $2s^22p^63s^23p^63d^{10}nln'l'~J^\Pi$, with $nln'l'=4s^2/4s4p/4s5s$ (following the considered state).

(2) Keep the orbitals fixed from step (1), and optimize an orbital basis layer by layer up to $nl=9h$ described by CSFs with the $J^\Pi$ symmetry of the state. These CSFs are obtained by SrD substitutions (at most one from the $2s^22p^63s^23p^63d^{10}$ core) on the SR set from step (1).

\vspace{-0.33cm}

\subsection{SrD-MR model}
\label{subsec:SrD-MR}

(1) Perform a calculation using an MR set consisting of CSFs with two forms: $2s^22p^63s^23p^63d^{10}nln'l'~J^\Pi$ with $nl,n'l'=4s,4p,4d,4f$ + $5s/\{5s,5p\}/\{5s,6s\}$, and $2s^22p^63s^23p^63d^9nln'l'n''l''~J^\Pi$ with $nl,n'l',n''l''=4s,4p,4d,4f$ + $5s/\{5s,5p\}/\{5s,6s\}$ (following the considered state). These CSFs account for a fair amount of the VV correlation, and for CV correlations between the $3d$ core orbital and the valence orbitals. \pagebreak

\onecolumngrid

\begin{table}[ht!]
\caption{\small{MR configurations for the lower and upper states of the two studied transitions in Zn~\textsc{i}. The MR-cutoff value, $\varepsilon_{\text{MR}}$, determines the set of CSFs in the MR space. $N_{\text{CSFs}}$ is the number of CSFs describing each MR space.}}
\vspace{0.1cm}
\begin{tabular}{l c c c l c l l c r}
\hline
\hline
\vspace{0.1cm}
Transition                                   & \hspace{0.5cm} & $\varepsilon_{\text{MR}}$ & \hspace{0.5cm} & $J^\Pi$ & \hspace{0.5cm} & \mc{2}{l}{MR configurations}                                                                                                   & \hspace{0.5cm} & $N_{\text{CSFs}}$ \\
\hline
$4s^{2}~^{1}S_{0} \rightarrow 4s4p~^{3}P^{o}_{1}$ & \hspace{0.5cm} & 0.01 & \hspace{0.5cm} & $0^+$  & \hspace{0.5cm} & [Ar]$3d^{10}\{4s^2,4p^2,4d^2\}$, & [Ar]$3d^{9}\{4s4p^2,4s4p4f,4s^24d\}$                     & \hspace{0.5cm} & 18 \\
\vspace{0.2cm}
                                                  & \hspace{0.5cm} &                                         & \hspace{0.5cm} & $1^-$   & \hspace{0.5cm} & [Ar]$3d^{10}\{4s4p,4p4d\}$,          & [Ar]$3d^{9}\{4s4p4d,4s4d4f,4p^3,4p^24f,4s^24p\}$ & \hspace{0.5cm} & 31 \\
$4s4p~^{3}P^{o}_{2} \rightarrow 4s5s~^{3}S_{1}$   & \hspace{0.5cm} & 0.01  & \hspace{0.5cm} & $2^-$   & \hspace{0.5cm} & [Ar]$3d^{10}\{4s4p,4p4d\}$,          & [Ar]$3d^{9}\{4s4p4d,4s4d4f,4p^3,4p^24f,4s^24p\}$ & \hspace{0.5cm} & 31 \\
\vspace{0.1cm}
                                                  & \hspace{0.5cm} &                                         & \hspace{0.5cm} & $1^+$  & \hspace{0.5cm} & [Ar]$3d^{10}\{4s5s,4p^2\}$,          & [Ar]$3d^{9}\{4s4d5s,4p^25s,4p4f5s\}$                      & \hspace{0.5cm} & 14 \\
\hline
\hline
\end{tabular}
\label{table_MR_composition}
\end{table}

\twocolumngrid

(2) Keep the orbitals fixed from step (1), and optimize an orbital basis layer by layer up to $nl=9h$ described by CSFs with the $J^\Pi$ symmetry of the state. These CSFs are obtained by SrD substitutions (at most one from the $2s^22p^63s^23p^63d^{10}$ core) on the MR set from step (1).

\vspace{-0.33cm}

\subsection{SrDT-SS model}
\label{subsec:SrDT-SS}

(1) Perform a calculation using a set consisting of CSFs with two forms: $2s^22p^63s^23p^63d^{9}nln'l'n''l''~J^\Pi$ and $2s^22p^63s^23p^53d^{10}nln'l'n''l''~J^\Pi$ with $nl,n'l',n''l''=4s,4p,4d,4f$ + $5s/\{5s,5p\}/\{5s,6s\}$ (following the considered state). These CSFs also account for a fair amount of the VV correlation, and for CV correlations between the $3p$ and $3d$ core orbitals and the valence orbitals. Add single $s$-substitutions (SS) by including the following CSFs: $2s^22p^63s3p^63d^{10}nln'l'n''l''~J^\Pi$ and $2s2p^63s^23p^63d^{10}nln'l'n''l''~J^\Pi$, with $nln'l'n''l''=4s^25s/4s4p5s/4s5s6s$.

(2) Keep the orbitals fixed from step (1), and optimize an orbital basis layer by layer up to $nl=9h$ described by CSFs with the $J^\Pi$ symmetry of the state. These CSFs are obtained by SrDT-SS substitutions (at most one from the $2s^22p^63s^23p^63d^{10}$ core) in the same way as in step (1). Although this model does not include all CV effects deep down in the core, it includes the ones that are important for getting accurate electron densities.

It is important to mention that core-core (CC) contributions, i.e., unrestricted SD substitutions from core orbitals, are not accounted for, contrary to the strategy adopted in the papers on Mg~\textsc{i}~\cite{FGE16} and Al~\textsc{i}~\cite{FBE16}. Indeed, the CSFs expansions in Zn~\textsc{i} become too large when CC correlations within the complete $2s^22p^63s^23p^63d^{10}$ core orbitals are added to the $nl=9h$ AS, counting for the $J^{\pi}=2^-$ state more than $10^8$~CSFs for the SrD-MR and SrDT-SS models. Such expansions exceed the capacity of our current computer resources by an order of magnitude. Restricting the CC correlations to only those within the $3d$ core orbital leads to around $10^7$~CSFs. Applying an SCF procedure takes too much computing time, but the use of the CI method would be feasible by means of a Brillouin-Wigner perturbative zero- and first order partition of the CSF space~\cite{KKT07,GJF17}. However, the computational task for estimating the IS factors with \textsc{ris{\small 3}} would exceed our current CPU time resources for such large expansions.

The CC correlation effects are known to be more balanced with a \textit{common} orbital basis for describing both upper and lower states, resulting in more accurate transition energies, as mentioned in Refs.~\cite{FGE16,FBE16,Ve87}. Hence, neglecting CC contributions enables us to use separate orbital basis sets, in which orbital relaxation is allowed.


\section{Numerical results}
\label{sec:results}

Let us first study the convergence of the level MS factors, $K_{\text{NMS}}$ and $K_{\text{SMS}}$ (in $m_eE_{\text{h}}$), and the electronic probability density at the origin, $\rho^e(\bs{0})$ (in $a_0^{-3}$), of a given transition as a function of the increasing AS. Tables{~\ref{table_KNMS_KSMS_rho1}} and {\ref{table_KNMS_KSMS_rho2}} display the SrD-SR, SrD-MR and SrDT-SS values respectively for the $4s^{2}~^{1}S_{0} \rightarrow 4s4p~^{3}P^{o}_{1}$ and $4s4p~^{3}P^{o}_{2} \rightarrow 4s5s~^{3}S_{1}$ transitions. The AS is extended until convergence of the differential results $\Delta^u_l$ is achieved, which requires the $nl=9h$ correlation layer (``CV~$9h$'').

Let us start the analysis with the $4s^{2}~^{1}S_{0} \rightarrow 4s4p~^{3}P^{o}_{1}$ transition. A satisfactory convergence is found for the three correlation models. The relative difference between the ``CV~$8h$'' and ``CV~$9h$'' values is $0.3-2.2\%$ for $\Delta K_{\text{NMS}}$, $0.8-2.5\%$ for $\Delta K_{\text{SMS}}$ and $0.4-0.5\%$ for $\Delta \rho^e(\bs{0})$, following the model. The analysis is similar for the $4s4p~^{3}P^{o}_{2} \rightarrow 4s5s~^{3}S_{1}$ transition, where the CV~$8h-$CV~$9h$ relative differences reach $0.3-1.1\%$ for $\Delta K_{\text{NMS}}$, $1.4-2.4\%$ for $\Delta K_{\text{SMS}}$ and $0.5-1.6\%$ for $\Delta \rho^e(\bs{0})$.

For both transitions, the relative differences are larger for $\Delta K_{\text{SMS}}$, as expected from the two-body nature of the SMS operator, which makes it more sensitive to electron correlation than the one-body NMS and density operators. However the convergence achieved for the SMS factors is highly satisfactory, remembering that small variations in the level values due to correlation effects can lead to a significant variation in the transition values. This illustrates the challenge of obtaining reliable values for the SMS factors with such a computational approach.

At this stage, convergence within the three correlation models has been investigated. However accuracy is not obviously implied, simply because the models may not be suitable for the studied properties. Hence, one also needs to compare the obtained results of the transition energies and IS factors with reference values. \tablename{~\ref{table_E}} displays the energies, $\Delta E$ (in cm$^{-1}$), of the two studied transitions in Zn~\textsc{i}. The SrD-SR, SrD-MR and SrDT-SS CV $9h$ values are compared with experimental NIST data~\cite{KRR15} and theoretical results~\cite{BG80,GM03,LHZ06,FZ07,CC10,LGZ11}. \pagebreak

\onecolumngrid

\begin{table}[ht!]
\caption{\small{Level MS factors, $K_{\text{NMS}}$ and $K_{\text{SMS}}$ (in $m_eE_{\text{h}}$), and electronic probability density at the origin, $\rho^e(\bs{0})$ (in $a_0^{-3}$), as functions of the increasing AS for the $4s^{2}~^{1}S_{0} \rightarrow 4s4p~^{3}P^{o}_{1}$ transition in Zn~\textsc{i}. SrD-SR, SrD-MR and SrDT-SS results are displayed. $\Delta^u_l$ stands for the difference between the values of the upper level and the lower level.}}
\vspace{0.1cm}
\begin{tabular}{l c c c c c c c c c c c c}
\hline
\hline
\vspace{0.1cm}
               & \hspace{0.5cm} & \mc{3}{c}{$K_{\text{NMS}}$ ($m_eE_{\text{h}}$)} & \hspace{0.5cm} & \mc{3}{c}{$K_{\text{SMS}}$ ($m_eE_{\text{h}}$)} & \hspace{0.5cm} & \mc{3}{c}{$\rho^e(\bs{0})$ ($a_0^{-3}$)} \\
\cline{3-5} \cline{7-9} \cline{11-13}
\vspace{0.1cm}
AS notation & \hspace{0.5cm}   & Lower             & Upper             & $\Delta^u_l$ & \hspace{0.5cm} & Lower            & Upper            & $\Delta^u_l$ & \hspace{0.5cm} & Lower               & Upper                & $\Delta^u_l$ \\
\hline
\mc{13}{c}{SrD-SR} \\
DHF         & \hspace{0.5cm} & $1779.7094$ & $1779.6087$ & $-0.1007$    & \hspace{0.5cm} & $-435.1390$ & $-435.3603$ & $-0.2213$    & \hspace{0.5cm} & $25020.8467$ & $25014.1784$ & $-6.6683$ \\
CV $4f$   & \hspace{0.5cm} & $1779.3810$ & $1779.3616$ & $-0.0194$    & \hspace{0.5cm} & $-434.1727$ & $-434.7012$ & $-0.5285$    & \hspace{0.5cm} & $25022.5672$ & $25015.3383$ & $-7.2289$ \\
CV $5g$  & \hspace{0.5cm} & $1779.4201$ & $1779.3906$ & $-0.0295$    & \hspace{0.5cm} & $-434.2497$ & $-434.7169$ & $-0.4672$    & \hspace{0.5cm} & $25022.7738$ & $25015.5299$ & $-7.2439$ \\
CV $6h$  & \hspace{0.5cm} & $1779.4854$ & $1779.4168$ & $-0.0686$    & \hspace{0.5cm} & $-434.2512$ & $-434.7035$ & $-0.4523$    & \hspace{0.5cm} & $25023.5160$ & $25015.8568$ & $-7.6592$ \\
CV $7h$  & \hspace{0.5cm} & $1779.4814$ & $1779.4152$ & $-0.0662$    & \hspace{0.5cm} & $-434.2538$ & $-434.6830$ & $-0.4292$    & \hspace{0.5cm} & $25023.4618$ & $25015.8565$ & $-7.6053$ \\
CV $8h$  & \hspace{0.5cm} & $1779.4882$ & $1779.4179$ & $-0.0703$    & \hspace{0.5cm} & $-434.2501$ & $-434.6797$ & $-0.4296$    & \hspace{0.5cm} & $25023.6107$ & $25015.9192$ & $-7.6915$ \\
\vspace{0.2cm}
CV $9h$  & \hspace{0.5cm} & $1779.4876$ & $1779.4179$ & $-0.0697$    & \hspace{0.5cm} & $-434.2531$ & $-434.6720$ & $-0.4189$    & \hspace{0.5cm} & $25023.5853$ & $25015.9275$ & $-7.6578$ \\
\mc{13}{c}{SrD-MR} \\
CV $4f$ (MR) & \hspace{0.5cm} & $1779.2665$ & $1779.3470$ & $~~0.0805$ & \hspace{0.5cm} & $-434.0901$ & $-434.7119$ & $-0.6218$ & \hspace{0.5cm} & $25022.5107$ & $25015.2195$ & $-7.2912$ \\
CV $5g$  & \hspace{0.5cm} & $1779.4103$ & $1779.3901$ & $-0.0202$    & \hspace{0.5cm} & $-434.1951$ & $-434.7086$ & $-0.5135$    & \hspace{0.5cm} & $25022.8213$ & $25015.5398$ & $-7.2815$ \\
CV $6h$  & \hspace{0.5cm} & $1779.4806$ & $1779.4195$ & $-0.0611$    & \hspace{0.5cm} & $-434.1945$ & $-434.6992$ & $-0.5047$    & \hspace{0.5cm} & $25023.5798$ & $25015.8522$ & $-7.7276$ \\
CV $7h$  & \hspace{0.5cm} & $1779.4787$ & $1779.4184$ & $-0.0603$    & \hspace{0.5cm} & $-434.1962$ & $-434.6776$ & $-0.4814$    & \hspace{0.5cm} & $25023.5231$ & $25015.8608$ & $-7.6623$ \\
CV $8h$  & \hspace{0.5cm} & $1779.4857$ & $1779.4213$ & $-0.0644$    & \hspace{0.5cm} & $-434.1917$ & $-434.6743$ & $-0.4826$    & \hspace{0.5cm} & $25023.6749$ & $25015.9234$ & $-7.7515$ \\
\vspace{0.2cm}
CV $9h$  & \hspace{0.5cm} & $1779.4861$ & $1779.4215$ & $-0.0646$    & \hspace{0.5cm} & $-434.1951$ & $-434.6668$ & $-0.4717$    & \hspace{0.5cm} & $25023.6442$ & $25015.9303$ & $-7.7139$ \\
\mc{13}{c}{SrDT-SS} \\
CV $4f$  & \hspace{0.5cm}  & $1779.3311$ & $1779.3772$ & $~~0.0461$ & \hspace{0.5cm} & $-434.1038$ & $-434.7129$ & $-0.6091$    & \hspace{0.5cm} & $25022.6098$ & $25015.3267$ & $-7.2831$ \\
CV $5g$  & \hspace{0.5cm} & $1779.4141$ & $1779.3852$ & $-0.0289$    & \hspace{0.5cm} & $-434.2114$ & $-434.7181$ & $-0.5067$    & \hspace{0.5cm} & $25022.8130$ & $25015.5123$ & $-7.3007$ \\
CV $6h$  & \hspace{0.5cm} & $1779.4690$ & $1779.4149$ & $-0.0541$    & \hspace{0.5cm} & $-434.2092$ & $-434.7107$ & $-0.5015$    & \hspace{0.5cm} & $25023.4879$ & $25015.8049$ & $-7.6830$ \\
CV $7h$  & \hspace{0.5cm} & $1779.4748$ & $1779.4133$ & $-0.0615$    & \hspace{0.5cm} & $-434.2096$ & $-434.6877$ & $-0.4781$    & \hspace{0.5cm} & $25023.4273$ & $25015.7942$ & $-7.6331$ \\
CV $8h$  & \hspace{0.5cm} & $1779.4835$ & $1779.4204$ & $-0.0631$    & \hspace{0.5cm} & $-434.2005$ & $-434.6866$ & $-0.4861$    & \hspace{0.5cm} & $25023.5226$ & $25015.8584$ & $-7.6642$ \\
\vspace{0.1cm}
CV $9h$  & \hspace{0.5cm} & $1779.4831$ & $1779.4186$ & $-0.0645$    & \hspace{0.5cm} & $-434.2020$ & $-434.6844$ & $-0.4824$    & \hspace{0.5cm} & $25023.4961$ & $25015.8391$ & $-7.6570$ \\
\hline
\hline
\end{tabular}
\label{table_KNMS_KSMS_rho1}
\end{table}

\begin{table}[ht!]
\caption{\small{Level MS factors, $K_{\text{NMS}}$ and $K_{\text{SMS}}$ (in $m_eE_{\text{h}}$), and electronic probability density at the origin, $\rho^e(\bs{0})$ (in $a_0^{-3}$), as functions of the increasing AS for the $4s4p~^{3}P^{o}_{2} \rightarrow 4s5s~^{3}S_{1}$ transition in Zn~\textsc{i}. SrD-SR, SrD-MR and SrDT-SS results are displayed. $\Delta^u_l$ stands for the difference between the values of the upper level and the lower level.}}
\vspace{0.1cm}
\begin{tabular}{l c c c c c c c c c c c c}
\hline
\hline
\vspace{0.1cm}
                & \hspace{0.5cm} & \mc{3}{c}{$K_{\text{NMS}}$ ($m_eE_{\text{h}}$)} & \hspace{0.5cm} & \mc{3}{c}{$K_{\text{SMS}}$ ($m_eE_{\text{h}}$)} & \hspace{0.5cm} & \mc{3}{c}{$\rho^e(\bs{0})$ ($a_0^{-3}$)} \\
\cline{3-5} \cline{7-9} \cline{11-13}
\vspace{0.1cm}
AS notation & \hspace{0.5cm}  & Lower              & Upper             & $\Delta^u_l$ & \hspace{0.5cm} & Lower            & Upper            & $\Delta^u_l$ & \hspace{0.5cm} & Lower               & Upper                & $\Delta^u_l$ \\
\hline
\mc{13}{c}{SrD-SR} \\
DHF         & \hspace{0.5cm} & $1779.6069$ & $1779.5190$ & $-0.0879$    & \hspace{0.5cm} & $-435.3606$ & $-435.2350$ & $0.1256$     & \hspace{0.5cm} & $25014.1581$ & $25016.5128$ & $2.3547$ \\
CV $4f$   & \hspace{0.5cm} & $1779.3680$ & $1779.3562$ & $-0.0118$    & \hspace{0.5cm} & $-434.7083$ & $-434.6908$ & $0.0175$     & \hspace{0.5cm} & $25015.3327$ & $25017.6406$ & $2.3079$ \\
CV $5g$  & \hspace{0.5cm} & $1779.3906$ & $1779.3685$ & $-0.0221$    & \hspace{0.5cm} & $-434.7230$ & $-434.6866$ & $0.0364$     & \hspace{0.5cm} & $25015.5156$ & $25017.8105$ & $2.2949$ \\
CV $6h$  & \hspace{0.5cm} & $1779.4165$ & $1779.3704$ & $-0.0461$    & \hspace{0.5cm} & $-434.7079$ & $-434.6412$ & $0.0667$     & \hspace{0.5cm} & $25015.8484$ & $25018.1404$ & $2.2920$ \\
CV $7h$  & \hspace{0.5cm} & $1779.4131$ & $1779.3759$ & $-0.0372$    & \hspace{0.5cm} & $-434.6870$ & $-434.6470$ & $0.0400$     & \hspace{0.5cm} & $25015.8434$ & $25018.1825$ & $2.3391$ \\
CV $8h$  & \hspace{0.5cm} & $1779.4152$ & $1779.3800$ & $-0.0352$    & \hspace{0.5cm} & $-434.6836$ & $-434.6517$ & $0.0319$     & \hspace{0.5cm} & $25015.9080$ & $25018.2378$ & $2.3298$ \\
\vspace{0.2cm}
CV $9h$  & \hspace{0.5cm} & $1779.4153$ & $1779.3800$ & $-0.0353$    & \hspace{0.5cm} & $-434.6765$ & $-434.6451$ & $0.0314$     & \hspace{0.5cm} & $25015.9164$ & $25018.2753$ & $2.3589$ \\
\mc{13}{c}{SrD-MR} \\
CV $4f$ (MR) & \hspace{0.5cm} & $1779.3530$ & $1779.3132$ & $-0.0398$ & \hspace{0.5cm} & $-434.7178$ & $-434.6881$ & $0.0297$  & \hspace{0.5cm} & $25015.2232$ & $25017.5319$ & $2.3087$ \\
CV $5g$  & \hspace{0.5cm} & $1779.3908$ & $1779.3614$ & $-0.0294$    & \hspace{0.5cm} & $-434.7132$ & $-434.6779$ & $0.0353$     & \hspace{0.5cm} & $25015.5309$ & $25017.8255$ & $2.2946$ \\
CV $6h$  & \hspace{0.5cm} & $1779.4181$ & $1779.3681$ & $-0.0500$    & \hspace{0.5cm} & $-434.7005$ & $-434.6348$ & $0.0657$     & \hspace{0.5cm} & $25015.8400$ & $25018.1319$ & $2.2919$ \\
CV $7h$  & \hspace{0.5cm} & $1779.4165$ & $1779.3794$ & $-0.0371$    & \hspace{0.5cm} & $-434.6794$ & $-434.6419$ & $0.0375$     & \hspace{0.5cm} & $25015.8486$ & $25018.2033$ & $2.3547$ \\
CV $8h$  & \hspace{0.5cm} & $1779.4193$ & $1779.3832$ & $-0.0361$    & \hspace{0.5cm} & $-434.6761$ & $-434.6464$ & $0.0297$     & \hspace{0.5cm} & $25015.9133$ & $25018.2413$ & $2.3280$ \\
\vspace{0.2cm}
CV $9h$  & \hspace{0.5cm} & $1779.4195$ & $1779.3833$ & $-0.0362$    & \hspace{0.5cm} & $-434.6694$ & $-434.6390$ & $0.0304$     & \hspace{0.5cm} & $25015.9190$ & $25018.2852$ & $2.3662$ \\
\mc{13}{c}{SrDT-SS} \\
CV $4f$  & \hspace{0.5cm}  & $1779.3812$ & $1779.3536$ & $-0.0276$    & \hspace{0.5cm} & $-434.7171$ & $-434.6938$ & $0.0233$     & \hspace{0.5cm} & $25015.3273$ & $25017.6190$ & $2.2917$ \\
CV $5g$  & \hspace{0.5cm} & $1779.3867$ & $1779.3616$ & $-0.0251$    & \hspace{0.5cm} & $-434.7232$ & $-434.6949$ & $0.0283$     & \hspace{0.5cm} & $25015.5067$ & $25017.7892$ & $2.2825$ \\
CV $6h$  & \hspace{0.5cm} & $1779.4137$ & $1779.3602$ & $-0.0535$    & \hspace{0.5cm} & $-434.7115$ & $-434.6473$ & $0.0642$     & \hspace{0.5cm} & $25015.7964$ & $25018.0507$ & $2.2543$ \\
CV $7h$  & \hspace{0.5cm} & $1779.4120$ & $1779.3770$ & $-0.0350$    & \hspace{0.5cm} & $-434.6903$ & $-434.6580$ & $0.0323$     & \hspace{0.5cm} & $25015.7839$ & $25018.1262$ & $2.3423$ \\
CV $8h$  & \hspace{0.5cm} & $1779.4183$ & $1779.3822$ & $-0.0361$    & \hspace{0.5cm} & $-434.6888$ & $-434.6538$ & $0.0350$     & \hspace{0.5cm} & $25015.8467$ & $25018.1656$ & $2.3189$ \\
\vspace{0.1cm}
CV $9h$  & \hspace{0.5cm} & $1779.4187$ & $1779.3822$ & $-0.0365$    & \hspace{0.5cm} & $-434.6816$ & $-434.6471$ & $0.0345$     & \hspace{0.5cm} & $25015.7979$ & $25018.1282$ & $2.3303$ \\
\hline
\hline
\end{tabular}
\label{table_KNMS_KSMS_rho2}
\end{table}

\twocolumngrid

\begin{table}[ht!]
\caption{\small{Energies, $\Delta E$ (in cm$^{-1}$), of the two studied transitions in Zn~\textsc{i}. Comparison with experimental NIST data~\cite{KRR15} and theoretical results~\cite{BG80,GM03,LHZ06,FZ07,CC10,LGZ11}. Values from Refs.~\cite{BG80,LGZ11} correspond to non-relativistic computations.}}
\vspace{0.1cm}
\begin{tabular}{l c c c c c c}
\hline
\hline
\mc{7}{c}{$\Delta E$ (cm$^{-1}$)} \\
\hline
\mc{3}{c}{This work}                               & \hspace{0.75cm} & NIST~\cite{KRR15} & \hspace{0.75cm} & Theory                            \\
\hline
\mc{7}{c}{$4s^{2}~^{1}S_{0} \rightarrow 4s4p~^{3}P^{o}_{1}$}                                                                                                 \\
SrD-SR   & \hspace{0.25cm} & $31\,878$ & \hspace{0.75cm} & $32\,501.421$    & \hspace{0.75cm} & $32\,153$~\cite{GM03}  \\
SrD-MR  & \hspace{0.25cm} & $32\,561$ & \hspace{0.75cm} &                             & \hspace{0.75cm} & $31\,804$~\cite{LHZ06} \\
SrDT-SS & \hspace{0.25cm} & $32\,460$ & \hspace{0.75cm} &                             & \hspace{0.75cm} & $32\,193$~\cite{FZ07}   \\
\vspace{0.2cm}
             & \hspace{0.25cm} &                  & \hspace{0.75cm} &                             & \hspace{0.75cm} & $32\,338$~\cite{CC10}   \\
\mc{7}{c}{$4s4p~^{3}P^{o}_{2} \rightarrow 4s5s~^{3}S_{1}$}                                                                                                     \\
SrD-SR   & \hspace{0.25cm} & $20\,769$ & \hspace{0.75cm} & $20\,781.928$    & \hspace{0.75cm} & $20\,547$~\cite{BG80}   \\
SrD-MR  & \hspace{0.25cm} & $20\,794$ & \hspace{0.75cm} &                             & \hspace{0.75cm} & $22\,488$~\cite{LGZ11} \\
\vspace{0.1cm}
SrDT-SS & \hspace{0.25cm} & $20\,732$ & \hspace{0.75cm} &                             & \hspace{0.75cm} &                                        \\
\hline
\hline
\end{tabular}
\label{table_E}
\end{table}

Let us consider the $4s^{2}~^{1}S_{0} \rightarrow 4s4p~^{3}P^{o}_{1}$ transition. G{\l}owacki and Migda{\l}ek~\cite{GM03} performed relativistic CI computations with Dirac-Fock wave functions using an \textit{ab initio} model potential. Liu \textit{et al.}~\cite{LHZ06} used the MCDHF method adopting a strategy on which the SrDT-SS model is based. Froese Fischer \textit{et al.}~\cite{FZ07} carried out MCHF and $B$-spline $R$-matrix calculations including Breit-Pauli corrections. Finally, Chen and Cheng~\cite{CC10} used $B$-spline basis functions for large-scale relativistic CI computations including QED corrections. \tablename{~\ref{table_E}} shows that the SrD-SR model provides a relative error of $1.9\%$ in comparison with NIST data. Better agreement is found with the more elaborate SrD-MR ($0.2\%$) and SrDT-SS models ($0.1\%$). It is clear from the comparison with the four above-cited theoretical works that our SrD-MR and SrDT-SS results show better agreement with NIST data.

In contrast to the $4s^{2}~^{1}S_{0} \rightarrow 4s4p~^{3}P^{o}_{1}$ transition, very few papers investigated the $4s4p~^{3}P^{o}_{2} \rightarrow 4s5s~^{3}S_{1}$ transition. To our knowledge, the only existing theoretical works were led by Bi{\'e}mont and Godefroid~\cite{BG80} using the MCHF method and by Liu \textit{et al.}~\cite{LGZ11} using the $R$-matrix method in the $LS$-coupling scheme. Both works are non-relativistic, and the transition energies must be compared with the $J$-averaged value $\Delta E=20\,975.905$ cm$^{-1}$ from NIST. \tablename{~\ref{table_E}} shows that the SrD-SR model provides a relative error of $0.06\%$ in comparison with NIST data, while the SrD-MR and SrDT-SS models respectively provide $0.06\%$ and $0.24\%$. Excellent agreement is thus found for all three models, and correlation beyond the SrD-SR model does not improve the accuracy on $\Delta E$.

Let us now compare the computed \textit{ab initio} IS electronic factors with reference results from the literature. As pointed out in Sec.~\ref{sec:intro}, most theoretical works report on properties in Zn~\textsc{i} and Zn-like ions other than IS factors. The only existing papers discussing SMS factors in Zn~\textsc{i} are seminal works in which low associated confidence is shown, compared with the accuracy to which IS measurements can be made~\cite{CBG97}. In addition, high-precision study of ISs has been carried out in the Zn$^+$ ion (Zn~\textsc{ii}). Kloch \textit{et al.}~\cite{KLS82} published measurements of optical ISs in the stable $^{64,66-68,70}$Zn isotopes for the $3d^{10}4p~^{2}P^{o}_{1/2} \rightarrow 3d^{9}4s^{2}~^{2}D_{3/2}$ (589.4 nm) transition in Zn~\textsc{ii}. Foot \textit{et al.}~\cite{FSS82} interpreted these measurements in terms of variations in the nuclear charge distribution. The measured ISs were separated into MS and FS contributions by combining the data with $\delta \langle r^2 \rangle$ results from electron scattering and muonic ($\mu$-e) IS experiments performed by Wohlfahrt \textit{et al.}~\cite{WSF80}. \pagebreak

Campbell \textit{et al.}~\cite{CBG97} measured ISs between the same stable isotopes for the $4s^{2}~^{1}S_{0} \rightarrow 4s4p~^{3}P^{o}_{1}$ (307.6 nm) transition. The ratio of FS factors, $F_{589.4}/F_{307.6}=-3.06(16)$, was extracted from a King plot using the IS measurements from Refs.~\cite{KLS82,FSS82}. The $F$-factor calculations of Blundell \textit{et al.}~\cite{BBP85,BBP87} enabled an estimate of $F_{307.6}=-1260$ MHz/fm$^{2}$ to be made. Note that the original value of $-1510$ MHz/fm$^{2}$ appearing in \Ref{CBG97} is actually a misprint~\cite{Ca16}.

Finally, the separation of the MS contribution proceeded through a King plot using the \textit{corrected} $F_{307.6}$ value together with $\delta \langle r^2 \rangle_{\mu-\text{e}}$ data from \Ref{WSF80}. Dividing the obtained MS between $^{66}$Zn and $^{64}$Zn isotopes, $\delta \nu_{\text{MS}}^{66,64}=921(31)$ MHz, by $(1/M_{66}-1/M_{64})$ yields $\Delta \tilde{K}_{\text{MS}}=-1970(29)$ GHz~u. The nuclear masses $M_{66}$ and $M_{64}$ are calculated by subtracting the mass of the electrons from the atomic masses, and by adding the binding energy~\cite{HAC76,LPT03,CST12}.

Yang \textit{et al.}~\cite{YWX16} measured ISs between the same stable isotopes for the $4s4p~^{3}P^{o}_{2} \rightarrow 4s5s~^{3}S_{1}$ (481.2 nm) transition. To calibrate the FS factor, a King plot was made using their set of ISs against the measured ISs from \Ref{CBG97}. This process enabled an estimate of $F_{481.2}=301(51)$ MHz/fm$^{2}$ to be made, assuming an error of $10\%$ on the \textit{erroneous} $F_{307.6}$ value of $-1510$ MHz/fm$^{2}$. \pagebreak

\onecolumngrid

\begin{table}[ht!]
\caption{\small{MS factors, $\Delta \tilde{K}_{\text{NMS}}$, $\Delta \tilde{K}_{\text{SMS}}$, and $\Delta \tilde{K}_{\text{MS}}$ (in GHz~u), and FS factors, $F$ (in MHz/fm$^{2}$), of the two studied transitions in Zn~\textsc{i}. Comparison of $\Delta \tilde{K}_{\text{NMS}}$ with values from the scaling law~\rref{eq_sl1}, and of $\Delta \tilde{K}_{\text{SMS}}$ and $F$ with results from Refs.~\cite{CBG97,YWX16,Ya16}.}}
\vspace{0.1cm}
\begin{tabular}{l c c c c c c l c c l c c l}
\hline
\hline
\mc{5}{c}{$\Delta \tilde{K}_{\text{NMS}}$ (GHz~u)}                 & \hspace{0.5cm} & \mc{2}{c}{$\Delta \tilde{K}_{\text{SMS}}$ (GHz~u)} & \hspace{0.5cm} & \mc{2}{c}{$\Delta \tilde{K}_{\text{MS}}$ (GHz~u)} & \hspace{0.5cm} & \mc{2}{c}{$F$ (MHz/fm$^{2}$)}               \\
\cline{1-5} \cline{7-8} \cline{10-11} \cline{13-14}
\mc{3}{c}{This work}   & \hspace{0.25cm} & Scal.~\rref{eq_sl1} & \hspace{0.5cm} & This work & \mc{1}{c}{Other}                                    & \hspace{0.5cm} & This work  & \mc{1}{c}{Other}                                 & \hspace{0.5cm} & This work & \mc{1}{c}{Other}                     \\
\hline
\mc{14}{c}{$4s^{2}~^{1}S_{0} \rightarrow 4s4p~^{3}P^{o}_{1}$} \\
SrD-SR  & & $-252$     & \hspace{0.25cm} & $-535$                 & \hspace{0.5cm}  & $-1512$  & $-1435(29)$~\cite{CBG97}                    & \hspace{0.5cm} & $-1764$   & $-1970(29)$~\cite{CBG97}                 & \hspace{0.5cm} & $-1131$  & $-1260$~\cite{CBG97}             \\
SrD-MR & & $-233$      & \hspace{0.25cm} &                             & \hspace{0.5cm}  & $-1703$  &                                                              & \hspace{0.5cm} & $-1936$   &                                                            & \hspace{0.5cm} & $-1139$  &                                                 \\
\vspace{0.2cm}
SrDT-SS & & $-233$    & \hspace{0.25cm} &                             & \hspace{0.5cm}  & $-1741$  &                                                              & \hspace{0.5cm} & $-1974$    &                                                            & \hspace{0.5cm} & $-1130$  &                                                 \\
\mc{14}{c}{$4s4p~^{3}P^{o}_{2} \rightarrow 4s5s~^{3}S_{1}$} \\
SrD-SR  & & $-127$     & \hspace{0.25cm} & $-342$                 & \hspace{0.5cm}  & $~113$   & $269(15)$~\cite{YWX16,Ya16}              & \hspace{0.5cm} & $~-14$     & $-73(15)$~\cite{YWX16,Ya16}             & \hspace{0.5cm} & $~348$   & $251(42)$~\cite{YWX16,Ya16} \\
SrD-MR  & & $-131$     & \hspace{0.25cm} &                             & \hspace{0.5cm}  & $~110$   &                                                              & \hspace{0.5cm} & $~-21$     &                                                             & \hspace{0.5cm} & $~349$   &                                                 \\
\vspace{0.1cm}
SrDT-SS & & $-132$    & \hspace{0.25cm} &                             & \hspace{0.5cm}  & $~125$    &                                                              & \hspace{0.5cm} & $-7$         &                                                             & \hspace{0.5cm} & $~343$   &                                                 \\
\hline
\hline
\end{tabular}
\label{table_KNMS_KSMS_F}
\end{table}

\twocolumngrid

To calibrate the MS factor, another King plot involving their set of ISs together with the calibrated $F_{481.2}$ value and $\delta \langle r^2 \rangle_{\mu-\text{e}}$ data from \Ref{WSF80} enabled the extraction of $\Delta \tilde{K}_{\text{MS}}=-59(18)$ GHz~u, adopting the sign conventions \rref{eq_delta_nu_MS} and \rref{eq_delta_nu_FS} of the present work. After correction of the $F_{307.6}$ value from $-1510$ MHz/fm$^{2}$ to $-1260$ MHz/fm$^{2}$, the FS and MS factors become $F_{481.2}=251(42)$ MHz/fm$^{2}$ and $\Delta \tilde{K}_{\text{MS}}=-73(15)$ GHz~u~\cite{Ya16}.

\begin{figure}[ht!]
\begin{minipage}[c]{.49\textwidth}
\includegraphics[scale=0.38]{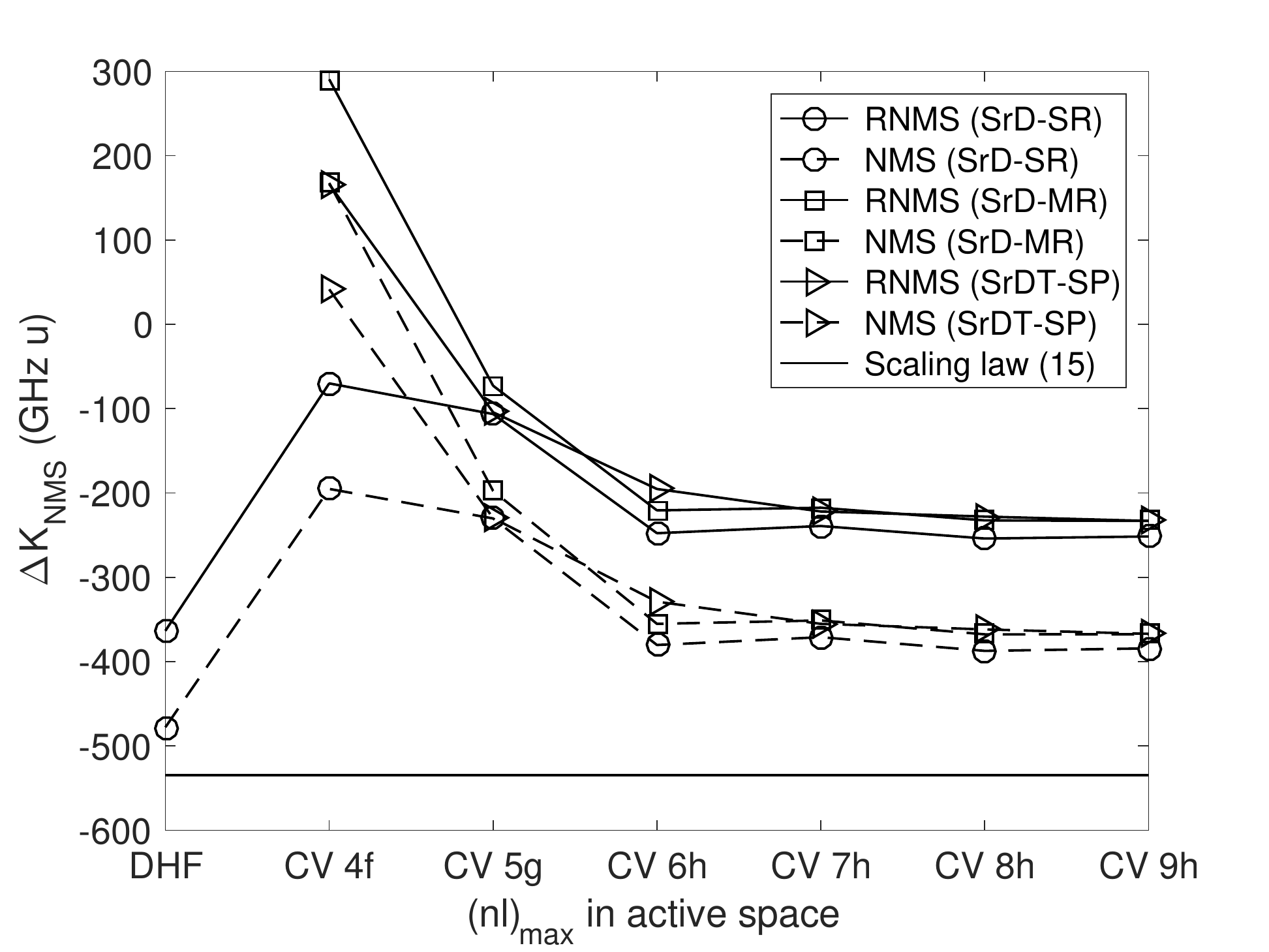}
\vspace{-0.1cm}
\begin{center}
(a)
\end{center}
\end{minipage}
\begin{minipage}[c]{.49\textwidth}
\includegraphics[scale=0.38]{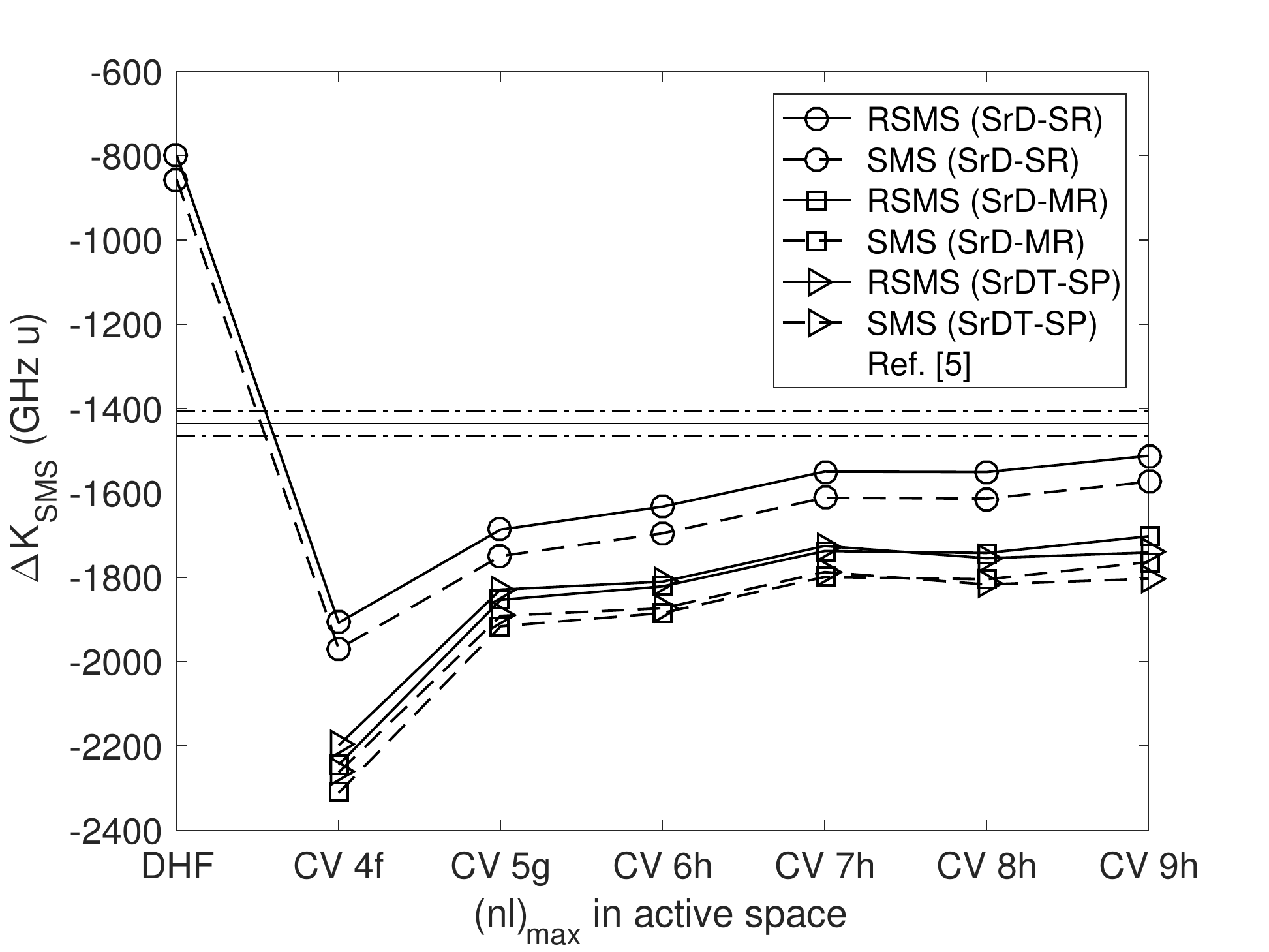}
\vspace{-0.1cm}
\begin{center}
(b)
\end{center}
\end{minipage}

\begin{minipage}[c]{.49\textwidth}
\includegraphics[scale=0.38]{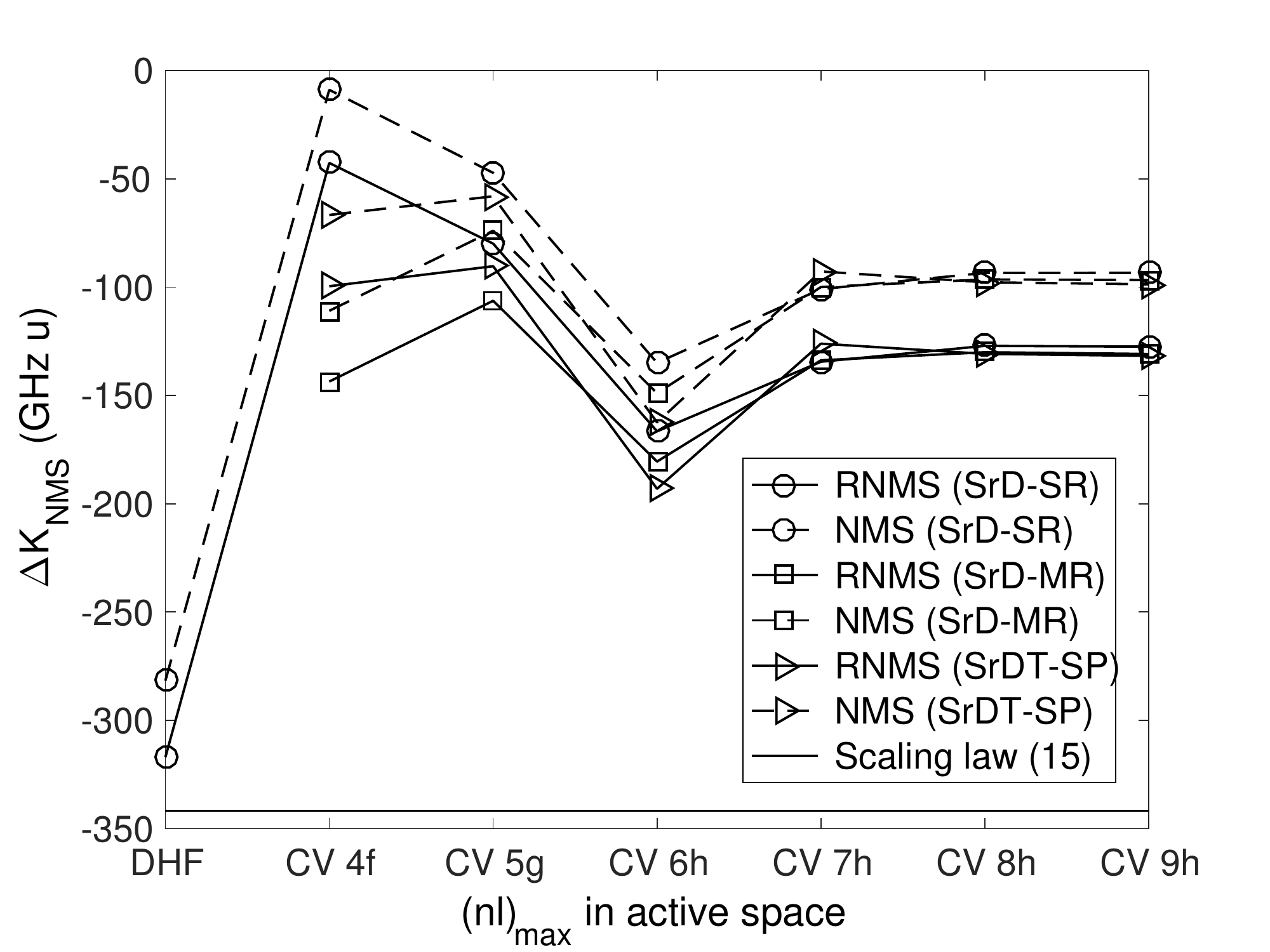}
\vspace{-0.1cm}
\begin{center}
(c)
\end{center}
\end{minipage}
\begin{minipage}[c]{.49\textwidth}
\includegraphics[scale=0.38]{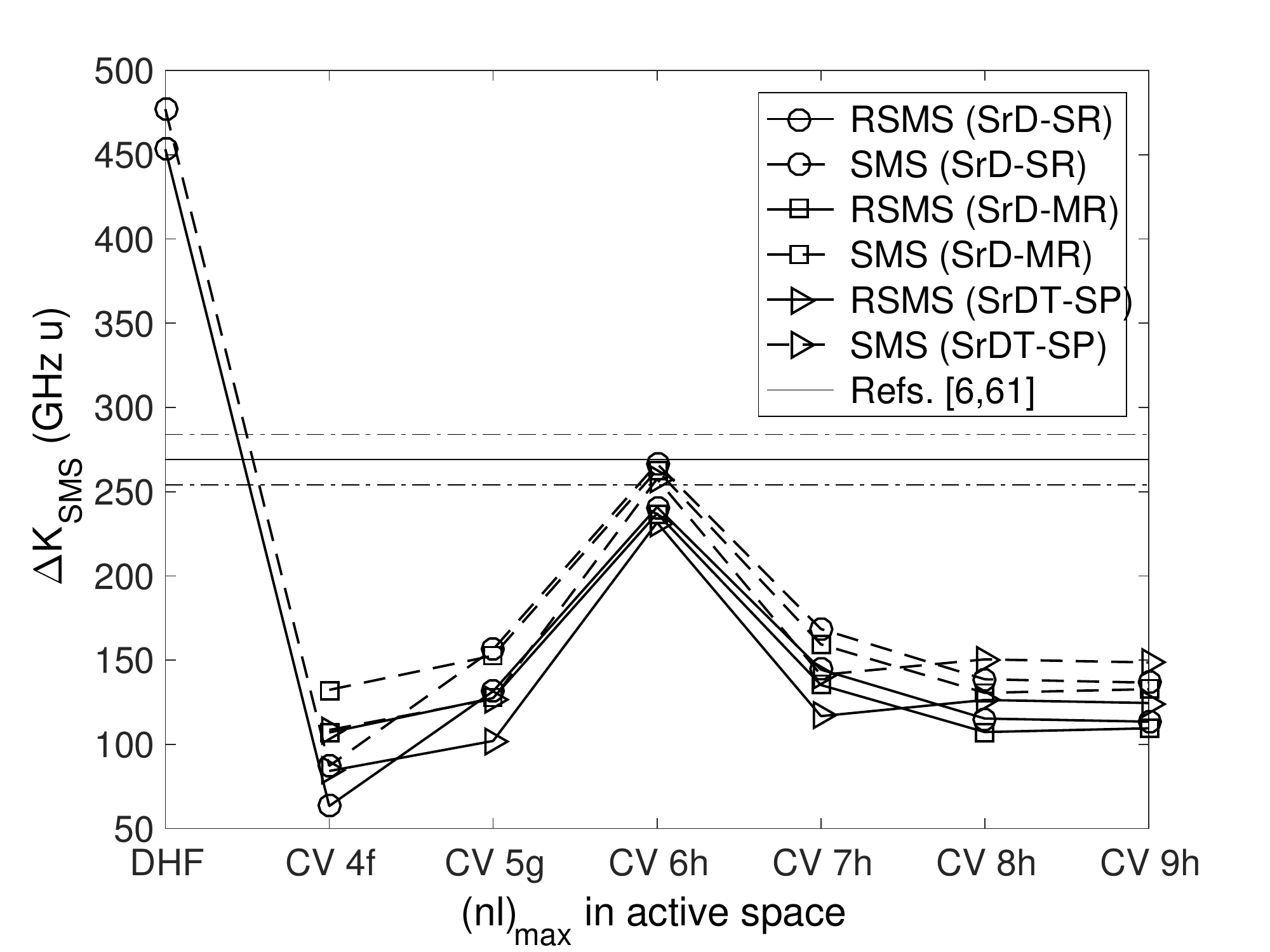}
\vspace{-0.1cm}
\begin{center}
(d)
\end{center}
\end{minipage}
\caption{\small{MS factors, $\Delta \tilde{K}_{\text{NMS}}$ and $\Delta \tilde{K}_{\text{SMS}}$ (in GHz~u), as functions of the increasing AS for (a)$-$(b) the $4s^2~^1S_0 \rightarrow 4s4p~^3P^o_1$ transition and (c)$-$(d) the $4s4p~^3P^o_2 \rightarrow 4s5s~^3S_1$ transition in Zn~\textsc{i}. ``RNMS'' and ``RSMS'' labels (solid lines) refer to the expectation values of the relativistic recoil Hamiltonian~\rref{eq_H_MS} while ``NMS'' and ``SMS'' labels (dashed lines) refer to the expectation values of its non-relativistic counterpart. $\Delta \tilde{K}_{\text{NMS}}$ are compared with values from the scaling law~\rref{eq_sl1}, and $\Delta \tilde{K}_{\text{SMS}}$ with results from Refs.~\cite{CBG97,YWX16,Ya16}. The horizontal dashed-dotted lines in (b) and (d) correspond to experimental uncertainties on $\Delta \tilde{K}_{\text{SMS}}$.}}
\label{fig_KNMS_KSMS}
\end{figure}

Experimentalists often split the total MS into the NMS and SMS contributions by estimating the NMS factor, $\Delta \tilde{K}_{k,\text{NMS}}$, with the scaling law approximation as
\beq
\Delta \tilde{K}_{k,\text{NMS}} \approx -m_e \nu_k^{\text{expt}},
\eeqn{eq_sl1}
where $m_e$ is the mass of the electron and $\nu_k^{\text{expt}}$ is the experimental transition energy of transition $k$, available in the NIST database~\cite{KRR15}. Doing so, one obtains $\Delta \tilde{K}_{\text{NMS}}=-535$ GHz~u for the $4s^{2}~^{1}S_{0} \rightarrow 4s4p~^{3}P^{o}_{1}$ transition and $\Delta \tilde{K}_{\text{NMS}}=-342$ GHz~u for the $4s4p~^{3}P^{o}_{2} \rightarrow 4s5s~^{3}S_{1}$ transition, respectively yielding the SMS contributions $\Delta \tilde{K}_{\text{SMS}}=-1435(29)$ GHz~u and $\Delta \tilde{K}_{\text{SMS}}=269(15)$ GHz~u. \tablename{~\ref{table_KNMS_KSMS_F}} displays the SrD-SR, SrD-MR and SrDT-SS CV~$9h$ MS factors, $\Delta \tilde{K}_{\text{NMS}}$, $\Delta \tilde{K}_{\text{SMS}}$, and $\Delta \tilde{K}_{\text{MS}}$ (in GHz~u), and FS factors, $F$ (in MHz/fm$^{2}$), of the two studied transitions in Zn~\textsc{i}. The values of $\Delta \tilde{K}_{\text{NMS}}$ are compared with the results from \Eq{eq_sl1} (``Scal.''), those of $\Delta \tilde{K}_{\text{SMS}}$ and $F$ with results from Refs.~\cite{CBG97,YWX16,Ya16}. Equation~\rref{eq_sl1} is only strictly valid in the non-relativistic framework, and the relativistic nuclear recoil corrections to $\Delta \tilde{K}_{\text{NMS}}$ can be computed with \textsc{ris\small{3}} as the expectation values of the relativistic part of the one-body term in the nuclear recoil Hamiltonian~\rref{eq_H_MS}, as shown in \figurename{~\ref{fig_KNMS_KSMS}}.

Let us start the comparison of the IS factors with the $4s^2~^1S_0 \rightarrow 4s4p~^3P^o_1$ transition. After correction, the FS factor from \Ref{CBG97} is in better agreement with our values, the relative difference reaching $10\%$. Moreover, the three models provide values in the same range, as expected from the one-body nature of the density operator. Turning to the total MS factor, it is seen that $\Delta \tilde{K}_{\text{MS}}$ is in excellent agreement with \Ref{CBG97} for the SrDT-SS model while it does not agree within the experimental error bars for the SrD-MR model, although the discrepancies are not large. By contrast, the SrD-SR model provides a number $200$ GHz~u higher, illustrating the sensitivity to electron correlation of the two-body SMS operator. Hence, correlation beyond the SrD-SR model improves the accuracy on $\Delta \tilde{K}_{\text{MS}}$.

Analysing the NMS and SMS factors separately, \tablename{~\ref{table_KNMS_KSMS_F}} shows that the three models provide $\Delta \tilde{K}_{\text{NMS}}$ values in the same range, as for the FS factor. Moreover, these results totally disagree with the number from the scaling law. \figurename{~\ref{fig_KNMS_KSMS}} shows that the relativistic nuclear recoil corrections to $\Delta \tilde{K}_{\text{NMS}}$ are important ($134$ GHz~u), representing around $+33\%$ of the NMS results obtained when neglecting them. The extracted $\Delta \tilde{K}_{\text{SMS}}$ value is also in disagreement with the results from the three models, as expected from the analysis of $\Delta \tilde{K}_{\text{NMS}}$. \figurename{~\ref{fig_KNMS_KSMS}} shows that the relativistic corrections to $\Delta \tilde{K}_{\text{SMS}}$ are much less important ($62$ GHz~u), representing around $+3\%$ of the SMS results obtained when neglecting them. In addition, these corrections on both MS factors are insensitive to electron correlation, staying constant along the increasing AS and being independent from the model.

It is shown from this analysis that only the sum of the NMS and SMS factors can be comparable with observation, the total $\mathcal{H}_{\text{MS}}$ being the only MS operator corresponding to an observable. The relativistic corrections partly cancel when summing $\Delta \tilde{K}_{\text{NMS}}$ and $\Delta \tilde{K}_{\text{SMS}}$, leading to $196$ GHz~u for the three models, which represent $10-11\%$ of the relativistic $\Delta \tilde{K}_{\text{MS}}$ values displayed in \tablename{~\ref{table_KNMS_KSMS_F}}. Hence, neglecting these corrections would bring the SrD-MR and SrDT-SS values around $200$ GHz~u too low in comparison with the experimental number.

The analysis is different for the $4s4p~^3P^o_2 \rightarrow 4s5s~^3S_1$ transition. None of the computed FS factors, whose average value is $F=346(3)$ MHz/fm$^2$, agrees with the number from Refs.~\cite{YWX16,Ya16}, although the three models provide values very close to each other. Turning to the total MS factor, an average of $\Delta \tilde{K}_{\text{MS}}=-14(7)$ GHz~u can be deduced from the three computed values. Thus, important discrepancy is found between this result and the number from Refs.~\cite{YWX16,Ya16}. \pagebreak

\onecolumngrid

\twocolumngrid

Moreover, it is not clear that correlation beyond the SrD-SR model improves the accuracy on $\Delta \tilde{K}_{\text{MS}}$ for this transition. Indeed, in contrast to the previous transition a strong cancellation is observed between the values of the NMS and SMS factors. Hence, small variations of the SMS factor due to correlation effects can significantly influence the total MS factor, leading to large theoretical error bars on the latter when comparing the models.

Analysing the NMS and SMS factors separately, \tablename{~\ref{table_KNMS_KSMS_F}} shows that the three models provide $\Delta \tilde{K}_{\text{NMS}}$ values in the same range, as for the previous transition. Again, these results totally disagree with the number from the scaling law. \figurename{~\ref{fig_KNMS_KSMS}} shows that the relativistic corrections to $\Delta \tilde{K}_{\text{NMS}}$ remain as important as in the previous transition ($-34$ GHz~u), representing around $-33\%$ of the NMS results obtained when neglecting them. The extracted $\Delta \tilde{K}_{\text{SMS}}$ value is also in disagreement with our results. \figurename{~\ref{fig_KNMS_KSMS}} shows that the relativistic corrections to $\Delta \tilde{K}_{\text{SMS}}$ are twice less important than for $\Delta \tilde{K}_{\text{NMS}}$ ($-23$ GHz~u), representing around $-17\%$ of the SMS results obtained when neglecting them. In addition, these two corrections are also insensitive to electron correlation.

Again, one concludes from this analysis that only the total MS factor is likely to be comparable with observation, although the discrepancy between theory and experiment is much higher than for the first transition. When summing $\Delta \tilde{K}_{\text{NMS}}$ and $\Delta \tilde{K}_{\text{SMS}}$, the relativistic corrections reach $-57$ GHz~u for the three models, which represents more than twice the relativistic $\Delta \tilde{K}_{\text{MS}}$ values displayed in \tablename{~\ref{table_KNMS_KSMS_F}}. Hence, neglecting these corrections would change the sign of all three values, and the agreement with the experiment would be worse.

Finally, fully non-relativistic MCHF computations of SMS factors are carried out for the two transitions of interest, using the \textsc{atsp{\small 2}k} program package~\cite{FTG07} and following the computational strategy of the SrD-SR and SrDT-SS models. For the $4s^2~^1S_0 \rightarrow 4s4p~^3P^o_1$ transition, the CV $9h$ values are $\Delta \tilde{K}_{\text{SMS}}=-1511$ GHz~u for SrD-SR and $-1731$ GHz~u for SrDT-SS, in excellent agreement with the fully relativistic results displayed in \tablename{~\ref{table_KNMS_KSMS_F}}. Hence, the relativistic corrections to the wave functions counterbalance the relativistic corrections to the $\mathcal{H}_{\text{SMS}}$ operator for this transition. For the $4s4p~^3P^o_2 \rightarrow 4s5s~^3S_1$ transition, the CV $9h$ values are $\Delta \tilde{K}_{\text{SMS}}=178$ GHz~u for SrD-SR and $175$ GHz~u for SrDT-SS, around $50-60$ GHz~u higher than the relativistic results from \tablename{~\ref{table_KNMS_KSMS_F}}. Hence, the relativistic corrections to the wave functions add to those to the $\mathcal{H}_{\text{SMS}}$ operator for this transition.

Attempts to solve the discrepancies highlighted in this work are ongoing~\cite{Ya17}. Yang \textit{et al.} are reinvestigating the extraction of the MS factor for the $4s4p~^3P^o_2 \rightarrow 4s5s~^3S_1$ transition, using the present computed average value of $F_{481.2}=348(1)$ MHz/fm$^2$ (with an associated $10-15\%$ error) in several King plots, together with their $481.2$ nm ISs, the $589.4$ nm ISs from \Ref{FSS82} and $\delta \langle r^2 \rangle_{\mu-\text{e}}$ data from \Ref{WSF80}. Since it is shown that inconsistency occurs in both FS and MS factors when plotting the $481.2$ nm ISs against the $307.6$ nm ones, coauthors of~\cite{Ya17} try to calibrate the MS factor without using \Ref{CBG97}. Moreover, as the present $\Delta \tilde{K}_{\text{NMS}}$ values do not agree with the scaling law, the NMS factor will not be fixed in the fit processes, contrary to the procedure adopted in the previous calibration. Note that the actual aim of \Ref{Ya17} is the determination of accurate $\delta \langle r^2 \rangle$ values between the $^{64,66-68,70}$Zn stable isotopes, using the present computed $F_{481.2}$ factor and the new calibrated 481.2 nm MS factor.


\vspace{-0.25cm}

\section{Conclusions}
\label{sec:conc}

This work describes \textit{ab initio} relativistic calculations of IS electronic factors in many-electron atoms using the MCDHF approach. The adopted computational approach for the estimation of the MS and FS factors for two transitions between low-lying states in Zn~\textsc{i} is based on the expectation values of the relativistic recoil Hamiltonian for a given isotope, together with the FS factors estimated from the total electron densities at the origin. Three different correlation models are explored in a systematic way to determine a reliable computational strategy and estimate theoretical error bars of the IS factors.

Within each correlation model, the convergence of the level MS factors and the electronic probability density at the origin, as a function of the increasing active space, is studied for the $4s^2~^1S_0 \rightarrow 4s4p~^3P^o_1$ and $4s4p~^3P^o_2 \rightarrow 4s5s~^3S_1$ transitions. Satisfactory convergence is found within the three correlation models, and for both studied transitions. It is shown that small variations in the level values due to correlation effects can lead to more significant variations in the transition values, concerning mainly the SMS factors.

The accuracy of the results obtained from the different correlation models is investigated by comparison with reference values. The transition energies show good agreement with observation available in the NIST database. Moreover, for both transitions the $\Delta E$ results are more accurate than numbers provided by other theoretical works. Since most of these works report on properties other than IS factors, results obtained in the present work are compared with numbers extracted from two experiments. Good agreement of the computed FS and total MS factors is found for the $4s^2~^1S_0 \rightarrow 4s4p~^3P^o_1$ transition. By contrast, the results are not consistent with values extracted from two King-plot processes for the $4s4p~^3P^o_2 \rightarrow 4s5s~^3S_1$ transition.

Significant discrepancies between theory and experiment appear when using the scaling law approximation~\rref{eq_sl1} to separate the NMS from the total MS. Indeed, the $\Delta \tilde{K}_{\text{NMS}}$ results completely disagree with numbers provided by this approximation, illustrating the rather fast breakdown of this law based on non-relativistic theory with respect to the atomic number $Z$. This breakdown has already been highlighted in heavier systems~\cite{LNG12,PQB16}. In consequence, the $\Delta \tilde{K}_{\text{SMS}}$ results also disagree with the extracted experimental values. To investigate these discrepancies, relativistic nuclear recoil corrections to $\Delta \tilde{K}_{\text{NMS}}$ and $\Delta \tilde{K}_{\text{SMS}}$ are discussed and quantified for both transitions. It is shown that neglecting them leads to larger discrepancies with observation for the total $\Delta \tilde{K}_{\text{MS}}$ values. Finally, fully non-relativistic calculations of the SMS factors are carried out with the MCHF method, considering the SrD-SR and SrDT-SS models. It is shown that the relativistic corrections to the wave functions  counterbalance the relativistic nuclear recoil corrections for the $4s^2~^1S_0 \rightarrow 4s4p~^3P^o_1$ transition, while they add for the $4s4p~^3P^o_2 \rightarrow 4s5s~^3S_1$ transition.

From the theoretical point of view, it would be worthwhile to study the effects of the omitted CC correlations within the $3d$ core orbital. Considerable code development is necessary in order to perform such large calculations in a reasonable time. A common optimization of the orbital sets is also required. Another possible way to improve the accuracy of the present results is the use of the partitioned correlation function interaction (PCFI) approach~\cite{VRJ13}. It is based on the idea of relaxing the orthonormality restriction on the orbital basis, and breaking down the very large calculations in the traditional multiconfiguration methods into a series of smaller parallel calculations. This method is very flexible for targeting different electron correlation effects. Additionally, electron correlation effects beyond the SrD-MR and SrDT-SS models (such as quadruple substitutions) can be included perturbatively. Work is being done in these directions.


\vspace{-0.5cm}

\begin{acknowledgments}
This work has been partially supported by the Belgian F.R.S.-FNRS Fonds de la Recherche Scientifique (CDR J.0047.16), the BriX IAP Research Program No. P7/12 (Belgium). L.F. acknowledges the support from the FRIA. J.B. acknowledges financial support of the European Regional Development Fund in the framework of the Polish Innovation Economy Operational Program (contract no. POIG.02.01.00-12-023/08). P.J. acknowledges financial support from the Swedish Research Council (VR), under contract 2015-04842.
\end{acknowledgments}


\end{document}